# Epistemological approach in immersive virtual environments and the neurophysiology learning process

Cesar R. Salas Guerra
Universitat Autónoma de Barcelona
P.O Box 931, Hatillo, PR 00659, USA
cesar.salasg@e-campus.uab.cat

## ABSTRACT
Currently virtual reality (VR) usage in training processes is increasing due to their usefulness in the learning processes based on visual information empowered. The information in virtual environments is perceived by sight, sound and touch, but the relationship or impact that these stimuli can have on the oscillatory activity of the brain such as the processing, propagation and synchronization of information still needs to be established in relation to the cognitive load of attention. Therefore, this study seeks to identify the suggested epistemological basis through literature review and current research agendas in the relationship that exists between the immersive virtual environment and the neurophysiology of learning processes by means of the analysis of visual information. The suggested dimensional modeling of this research is composed by the theory of information processing which allows the incorporation of learning through stimuli with the use of attention, perception and storage by means of information management and the Kolb's learning model which defines the perception and processing of information as dimensions of learning. Regarding to the neurophysiology of learning, the literature has established he links between the prefrontal cortex and working memory within the process of information management. The challenges and advances discussed in this research are based in the relationship between the identified constructs (Income Stimuli, Information Management and Cognitive Processing) and the establishment of a research agenda on how to identify the necessary indicators to measure memory and attention in the virtual immersion environments.

## 1. INTRODUCTION
Currently augmented reality (AR) and virtual reality (VR) usage in consumer-oriented technologies (Bonetti, Warnaby, & Quinn, 2018) and training processes (Andersen, Mikkelsen, Konge, Cayé-Thomasen, & Sørensen, 2016) is increasing due to their usefulness for learning processes based on visual information empowered by its communication and information presentation capabilities (Seeber et al., 2018).

Immersive virtual environments enable the manipulation of visual information in different situations through the use of entities such as avatars, virtual humans (Diemer, Alpers, Peperkorn, Shiban, & Mühlberger, 2015) products for commercialization (Hilken et al., 2018) and services such as the training (Venturini et al., 2016) and cognitive processing (Cognifit, 2019) referring to a number of tasks the brain does continuously.

Among the new challenges in the development of multimodal presentations based on visual and audible stimuli, the integration of online experiences and their adaptation to personal environments have integrated omnichannel augmented reality experiences (Hilken et al., 2018) which merge digital content as images and animations in a physical environment where the user interacts in real time.

## 2. RESEARCH OBJETIVES
The information in virtual environments is perceived by sight, sound and touch, but the relationship or impact that these stimuli can have on the oscillatory activity of the brain such as the processing, propagation and synchronization of information (Kissinger, Pak, Tang, Masmanidis, & Chubykin, 2018) still needs to be established in relation to the cognitive load of attention.

Therefore, this study seeks to identify the suggested epistemological basis through literature review and current research agendas in relation to the immersive virtual environment and the neurophysiology of learning processes by means of the analysis of visual information.

## 3. REVIEW OF LITERATURE AND RESEARCH AGENDAS
Within the paradigm that establishes the potential of the virtual reality for revolutionizing education (Andersen et al., 2016) the current research agendas signal the lack of practical elements that solve the use of simulation in education (Jensen & Konradsen, 2018) as an effective method in learning.

However, for Seeber et al., (2018) adding information through audible and visual elements to the physical world through the production of an immersive environment can influence the capabilities of collaborative groups, virtual meetings and interaction spaces through communication and presentation of information.

Recent literature establishes that the biological system needs to make inferences about the environment in order to plan and achieve goals successfully, to make these inferences Młynarski & Hermundstad, (2018) propose that the nervous system can build an internal model that links incoming sensory stimuli to the behavioral properties.

Though, there are certain patterns of deficiency in visual function that the literature identifies as depth perception, peripheral vision, visual search, speed of visual processing and color perception (Navarra & Waterhouse, 2019) which are accompanied by a process in reduction of sensation to cognition linked in the frontal areas.

Similarly, recent studies have shown that the development of preparatory attentional processes presumes success in prospective tasks (Grandi & Tirapu-Ustárroz, 2017) since one of the qualities of attention is to help the extraction of the characteristics of a stimulus, increasing the brain activity involved in the processing thereof (Rothlein, DeGutis, & Esterman, 2016).

## 4. THEORETICAL BIAS
### 4.1 Information Processing Theory
The theory of information processing allows the incorporation of learning through stimuli with the use of attention, perception and storage (Moos, 2015) by means of information management. This allows the functioning of memory through the gradual increase of the information processing capacities in the development of skills for obtaining knowledge.

Consequently, the user executes cognitive coding processes that involve the processing of information in the memory (Andrews Acquah, 2017), so that it can be recovered later by acquiring information stored in it. This theory establishes that there are limits on the amount of information that is processed in each stage; this information processing being interactive and multitasking (Jack Snowman & Biehler, 2014).

### 4.2 Kolb's Learning Model
Not all people learn in the same way, therefore, it is important to use different learning styles, according to Hoffmann, Agustín Freiberg; Liporace, (2015) the styles are a series of cognitive, affective and physiological qualities, which allow us to know the way in which people perceive, respond and interact in different learning situations.

David Kolb defines the perception and processing of information as dimensions of learning (Rodríguez Cepeda, 2018), these highlight a certain level of importance in andragogic education, where this theorist states that the preference for abstract learning increases with age, decreasing this forms activity and pragmatism (Hoffmann, Agustín Freiberg; Liporace, 2015).

According to Romero Agudelo, Salinas Urbina, & Mortera Gutiérrez, (2010), this learning model incorporates four basic capacities: concrete experience (CE), reflexive observation (OR), abstract conceptualization (CA) and active experimentation (EA). Kolb states that experience has a relevance in the learning process (Rodríguez Cepeda, 2018) conditioning the life experience.

## 5. NEUROPHYSIOLOGY OF LEARNING AND INFORMATION MANAGEMENT
### 5.1 Sensory Function
Several functions have been attributed to the oscillatory activity in the brain, including the processing, propagation and synchronization of information (Kissinger et al., 2018). They correlate with cognitive load as attention, in the same way Einstein, Polack, Tran, & Golshani, (2017) state that low frequency oscillations are important within the processes of perception of visual stimuli.

The low-level visual properties have certain characteristics such as tone, spatial frequency and orientation (Ester, Sprague, & Serences, 2017), however, there are certain patterns of visual function deficiency that the literature identifies as the perception of depth, peripheral vision, visual search, visual processing speed and color perception (Navarra & Waterhouse, 2019).

These are accompanied by a process of reduction of sensation to cognition linked to the areas of frontal attention, therefore, recent studies have shown that differences in the space of perceptual similarity influence individual attention (Rothlein et al., 2016) modulating the degree of behavioral interference due to the similarity in the visual search task.

### 5.2 Memory Function
Memory performs a complex mental process through the coding, maintenance and retrieval of stored information (Parra-Bolaños, Fidel, & de la Peña, 2017), although recent studies have shown that damage to the temporal lobes can affect this capacity, the performance in short-term tasks is not affected (López, 2011).

Within the neurophysiology of mnemonic processes, literature has established over time he links between the prefrontal cortex and working memory (Santana, 2007). The latter is defined as a mechanism to store and use information available for a short period of time (Gontier B., 2011), and within the process of information management, the prospective memory has been conceptualized as an ability to remember and carry out certain operations in the future (Grandi & Tirapu-Ustárroz, 2017), this entails a process of information retrieval at the moment which is required of it (Fombuena, 2016).

Demonstrating its importance in the processes of coding and interpretation of stored information, understanding that short-term memories develop sufficient semantic recovery resources (Luna-Lario, Peña, & Ojeda, 2017). Although the iconic memory demands and requires attention (Parra-Bolaños et al., 2017) identifying in this way a relationship of this type of memory with attention.

### 5.3 Attention Function
The different neural networks distributed at the cortical and subcortical level, such as the reticular formation, amygdala, thalamus, basal ganglia, parietal cortex, frontal cortex and prefrontal cortex, are involved in care (Parra-Bolaños et al., 2017) however, this is actively involved in executive control and detection of stimuli.

Recent studies have shown that developing preparatory attentional processes suppose success in prospective tasks (Grandi & Tirapu-Ustárroz, 2017) since one of the qualities of attention is to help the extraction of the characteristics of a stimulus, thus increasing the brain activity involved in the processing of it (Ruiz-Contreras & Cansino, 2005).

Within the neurophysiological aspects, the neural networks involved in the attention, such as the warning or alert system, are responsible for maintaining the processes of receptivity to the stimuli, helping to prepare the response (Bartés-Serrallonga et al., 2014), located in the thalamus, which receives the information that comes from the senses.
Within the processes of attention, the relevance of the decision pre-supposes a concrete action determined in diverse neural influences (Santana, 2007), starting from the influence of the sensory areas of the posterior cortex in the motivations and instincts, which implies the ability to select weak but relevant stimuli, in stronger stimulus environments.

# 6. SUGGESTED THEORETICAL MODEL

This theoretical framework consists of three dimensions defined by references in relevant academic literature. This theoretical framework demonstrates an understanding of the constructs and their relation to the identified indicators. This analytical model is relevant to the research problem that is being investigated.

## 6.1 Model Dimensions

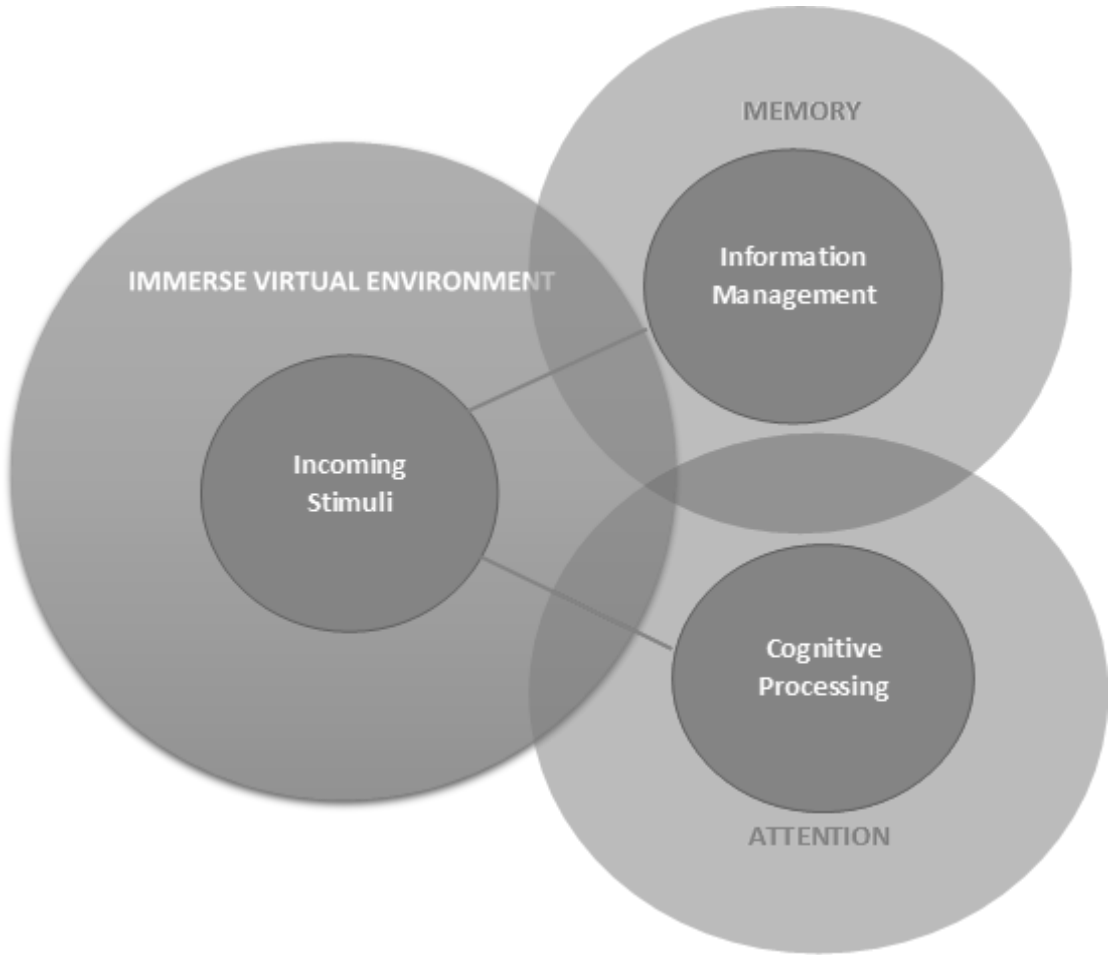

## 6.2 Variable Matrix

|  | ATN | STM | Reference |
|---|---|---|---|
| *Immerse Virtual Environment (IVE)* |  | Cognitive Processing | (Seeber et al., 2018) |
| *Memory (MEM)* | Information Management |  | (Parra-Bolaños, Fidel, & de la Peña, 2017) |
| *Attention (ATN)* |  | Cognitive Processing | (Grandi & Tirapu-Ustárroz, 2017) (Bartés-Serrallonga et al., 2014) |
| *Stimuli (STM)* | Information Management |  | (Kissinger, Pak, Tang, Masmanidis, & Chubykin, 2018) |
|  | *Construct* | *Construct* |  |

## 7. DISCUSSION AND ACADEMIC IMPLICATIONS

### 7.1 The Metatheory of Learning and Virtual Immersion Environments

The need for a revision arose in part from the problem of the vacuum that exists in this research topic. In addition, this motivation began by the literature analysis based in the theoretical evidence that integrates concepts and mechanisms in traditional learning within new virtual environment learning. The review of literature showed that this phenomenon is fertile ground for such attempts because several studies point to a relationship between stimuli and information processing through the use of memory, as well as the relationship of cognitive processes with attention.

The challenges and advances discussed so far relate to the findings found in the literature review that addresses the relationship between the identified constructs (Income Stimuli, Information Management and Cognitive Processing) and the establishment of a research agenda on how to identify the necessary indicators to measure memory and attention in the virtual immersion environments.

## 8. REFERENCES

[1] Andersen, S. A. W., Mikkelsen, P. T., Konge, L., Cayé-Thomasen, P., & Sørensen, M. S. (2016). The effect of implementing cognitive load theory-based design principles in virtual reality simulation training of surgical skills: a randomized controlled trial. Advances in Simulation, 1(1), 20. https://doi.org/10.1186/s41077-016-0022-1

[2] Andrews Acquah, E. Q. (2017). Using the Information Processing Approach to Explain the Mysteries of the Black Box : Implications for Teaching Religious and Moral Education. Journal of Information Engineering and Applications, 7(7), 1–4.

[3] Bartés-Serrallonga, M., Adan, A., Solé-Casals, J., Caldú, X., Falcón, C., Pérez-Pàmies, M., … Serra-Grabulosa, J. M. (2014). Bases cerebrales de la atención sostenida y la memoria de trabajo: Un estudio de resonancia magnética funcional basado en el continuous performance test. Revista de Neurologia, 58(7), 289–295.

[4] Bonetti, F., Warnaby, G., & Quinn, L. (2018). Augmented Reality and Virtual Reality in Physical and Online Retailing: A Review, Synthesis and Research Agenda. 119–132. https://doi.org/10.1007/978-3-319-64027-3_9

[5] Cognifit. (2019). Cognitive processes: What are they? Can they Improve and How? Retrieved July 10, 2019, from https://blog.cognifit.com/cognitive-processes/

[6] Diemer, J., Alpers, G. W., Peperkorn, H. M., Shiban, Y., & Mühlberger, A. (2015). The impact of perception and presence on emotional reactions: A review of research in virtual reality. Frontiers in Psychology, 6(JAN), 1–9. https://doi.org/10.3389/fpsyg.2015.00026

[7] Einstein, M. C., Polack, P.-O., Tran, D. T., & Golshani, P. (2017). Visually Evoked 3–5 Hz Membrane Potential Oscillations Reduce the Responsiveness of Visual Cortex Neurons in Awake Behaving Mice. The Journal of Neuroscience, 37(20), 5084–5098. https://doi.org/10.1523/jneurosci.3868-16.2017

[8] Ester, E. F., Sprague, T. C., & Serences, J. T. (2017). Category knowledge biases sensory representations in human visual cortex. BioRxiv, 025872, 170845. https://doi.org/10.1101/170845

[9] Fombuena, N. G. (2016). Normalización y validación de un test de memoria en envejecimiento normal, deterioro cognitivo leve y enfermedad de Alzheimer.

[10] Gontier B., J. (2011). Memoria de Trabajo y Envejecimiento. Revista de Psicología, 13(2), 111–124. https://doi.org/10.5354/0719-0581.2004.17804

[11] Grandi, F., & Tirapu-Ustárroz, J. (2017). Neuropsicología de la memoria prospectiva basada en el evento. Revista de Neurologia, 65(5), 226–233.

[12] Hilken, T., Heller, J., Chylinski, M., Keeling, D. I., Mahr, D., & de Ruyter, K. (2018). Making omnichannel an augmented reality: the current and future state of the art. Journal of Research in Interactive Marketing, 12(4), 509–523. https://doi.org/10.1108/JRIM-01-2018-0023

[13] Hoffmann, Agustín Freiberg; Liporace, M. M. F. (2015). ESTILOS DE APRENDIZAJE EN ESTUDIANTES UNIVERSITARIOS INGRESANTES Y AVANZADOS DE BUENOS AIRES 1 - ProQuest. Liberabit, 21(1), 71–79. https://doi.org/10.1364/JOSAB.23.002535

[14] Jack Snowman, & Biehler, R. (2014). Information Processing Theory. In Psychology Applied to Teaching (11th ed., pp. 1–21). Pennsylvania: Houghton Mifflin.

[15] Jensen, L., & Konradsen, F. (2018). A review of the use of virtual reality head-mounted displays in education and training. Education and Information Technologies, 23(4), 1515–1529. https://doi.org/10.1007/s10639-017-9676-0

[16] Kissinger, S. T., Pak, A., Tang, Y., Masmanidis, S. C., & Chubykin, A. A. (2018). Oscillatory Encoding of Visual Stimulus Familiarity. The Journal of Neuroscience, 38(27), 6223–6240. https://doi.org/10.1523/JNEUROSCI.3646-17.2018

[17] López, M. (2011). Memoria de trabajo y aprendizaje:aportes de la neuropsicologia. Cuad. Neuropsicolo, 5(1), 25–47. Retrieved from www.cnps.cl

[18] Luna-Lario, P., Peña, J., & Ojeda, N. (2017). Comparación de la escala de memoria de wechsler-iii y el test de aprendizaje verbal españa-complutense en el daño cerebral adquirido: Validez de constructo y validez ecológica. Revista de Neurologia, 64(8), 353–361.

[19] Młynarski, W. F., & Hermundstad, A. M. (2018). Adaptive coding for dynamic sensory inference. ELife, 7, 1–43. https://doi.org/10.7554/elife.32055

[20] Moos, D. (2015). Information Processing Theory in Context. In Educational Psychology. St. Peter, Minnesota.

[21] Navarra, R. L., & Waterhouse, B. D. (2019). Considering noradrenergically mediated facilitation of sensory signal processing as a component of psychostimulant-induced performance enhancement. Brain Research, 1709(October), 67–80. https://doi.org/10.1016/j.brainres.2018.06.027

[22] Parra-Bolaños, N., Fidel, M., & de la Peña, C. (2017). Atención y Memoria en estudiantes con bajo rendimiento académico. Un estudio exploratorio. Reidocrea, 6(7), 74–83. Retrieved from http://www.ugr.es/~reidocrea/6-7.pdf

[23] Rodríguez Cepeda, R. (2018). Los modelos de aprendizaje de Kolb, Honey y Mumford: implicaciones para la educación en ciencias. Sophia, 14(1), 73. https://doi.org/10.18634/sophiaj.14v.1i.698

[24] Romero Agudelo, L. N., Salinas Urbina, V., & Mortera Gutiérrez, F. J. (2010). Estilos de aprendizaje basados en el modelo de Kolb en la educación virtual. Apertura, 12(Abril), 72–85.

[25] Rothlein, D., DeGutis, J., & Esterman, M. (2016). Sensitivity to perceptual similarity is associated with greater sustained attention ability. Journal of Vision, 16(12), 420. https://doi.org/10.1167/16.12.420

[26] Ruiz-Contreras, A., & Cansino, S. (2005). Neurofisiología de la interacción entre la atención y la memoria episódica: revisión de estudios en modalidad visual. Revista de Neurologia, 41(12), 733–743.

[27] Santana, N. (2007). Receptores Monoaminérgicos en Corteza Prefrontal: Mecanismo de Acción de Fármacos Antipsicóticos. Universidad de Barcelona.

[28] Seeber, I., Bittner, E., Briggs, R. O., de Vreede, G.-J., de Vreede, T., Druckenmiller, D., … Söllner, M. (2018). Machines as Teammates: A Collaboration Research Agenda. Proceedings of the 51st Hawaii International Conference on System Sciences, 9, 420–429. https://doi.org/10.24251/hicss.2018.055

[29] Venturini, E., Riva, P., Serpetti, F., Romero Lauro, L., Pallavicini, F., Mantovani, F., … Parsons, T. D. (2016). A 3D virtual environment for empirical research on social pain: Enhancing fidelity and anthropomorphism in the study of feelings of ostracism inclusion and overinclusion. Annual Review of CyberTherapy and Telemedicine, 14, 89–94.